\documentclass[12pt]{article}

\usepackage{amsmath,amssymb}
\usepackage{epsfig,lscape,rotating}
\usepackage{graphicx}
\usepackage{color}

\def\beq{\begin{equation}}
\def\eeq#1{\label{#1}\end{equation}}
\def\eeqn{\end{equation}}
\def\beqa{\begin{eqnarray}}
\def\eeqa#1{\label{#1}\end{eqnarray}}
\def\eeqan{\end{eqnarray}}
\def\CR{\nonumber \\ }
\def\leqn#1{(\ref{#1})}

\def\stacksymbols #1#2#3#4{\def\theguybelow{#2}
    \def\vp{\lower#3pt}
    \def\sp{\baselineskip0pt\lineskip#4pt}
    \mathrel{\mathpalette\intermediary#1}}

\def\intermediary#1#2{\vp\vbox{\sp
     \everycr={}\tabskip0pt
     \halign{$\mathsurround0pt#1\hfil##\hfil$\crcr#2\crcr
              \theguybelow\crcr}}}

\def\gsim{\stacksymbols{>}{\sim}{2.5}{.2}}
\def\lsim{\stacksymbols{<}{\sim}{2.5}{.2}}

\def\to{\rightarrow}

\newcommand{\bspace}{\!\!\!\!}
\def\met{\mbox{$E{\bspace}/_{T}$}}

\def\n1{\tilde{\chi}^0_1}
\def\tq{U^\prime}

\begin{document}
\begin{titlepage}

\vskip.5cm
\begin{center}
{\huge \bf Model Discrimination at the LHC: \\ \vskip.5cm
a Case Study} \\
\vskip0.4cm
{\huge \bf } 
\vskip.2cm
\end{center}
\vskip1cm

\begin{center}
{\bf Gregory Hallenbeck, Maxim Perelstein, Christian Spethmann,\\ Julia 
Thom, and Jennifer Vaughan} \\
\end{center}
\vskip 8pt

\begin{center}
	{\it Newman Laboratory of Elementary Particle Physics\\
	Cornell University, Ithaca, NY 14853, USA } \\

\vspace*{0.3cm}

{\tt  glh59@cornell.edu, maxim@lepp.cornell.edu, cs366@cornell.edu,
jt297@cornell.edu, jjv27@cornell.edu}
\end{center}

\vglue 0.3truecm

\begin{abstract}
\vskip 3pt \noindent
We investigate the potential of the Compact Muon Solenoid (CMS) 
detector at the Large Hadron 
Collider (LHC) to discriminate between two theoretical models 
predicting anomalous events with jets and large missing transverse 
energy, minimal supersymmetry and Little Higgs with T Parity. We focus 
on a simple test case scenario, in which the only exotic particles 
produced at the LHC are heavy color-triplet states (squarks or 
T-quarks), and the only open decay channel for these particles is into 
the stable missing-energy particle (neutralino or heavy photon) plus a 
quark. We find that in this scenario, the angular and momentum 
distributions of the observed jets are sufficient to discriminate 
between the two models with a few inverse fb of the LHC data, provided 
that these distributions for both models and the dominant Standard 
Model backgrounds can be reliably predicted by Monte Carlo simulations.
\end{abstract}

\end{titlepage}

\section{Introduction}

Theoretical arguments strongly indicate that the Standard Model (SM) picture
of electroweak symmetry breaking is incomplete, and numerous  
extensions of the SM at the electroweak scale have been proposed. It is 
expected that at least some of the new particles and interactions predicted by 
such extended theories will be discovered and studied at the LHC. The 
ultimate goal of the experiments is, of course, to determine the correct 
theory of physics at the TeV scale. This task may be quite complicated. In
particular, it is quite likely that nature is described by one of the several 
models possessing the following features:

\begin{itemize}

\item Physics at the TeV scale is weakly coupled, and there is a light Higgs
(as motivated by precision electroweak data);

\item A number of new states are present at the TeV scale, and new particles 
can be paired up with the known SM states, with states in the same pair 
carrying identical gauge charges;

\item New states carry a parity quantum number distinct from their SM 
counterparts, implying that the lightest new particle (LNP) is stable;

\item The LNP is weakly interacting (as motivated by 
cosmological constraints on stable particles).

\end{itemize}

The best known model of this class is the minimal supersymmetric standard model
(MSSM). Other contenders include models with universal extra dimensions (UED) 
and Little Higgs models with T parity. Broadly speaking, all these 
theories share the same LHC phenomenology: the new physics production is 
dominated by the colored states, which are pair-produced, and then decay down 
to the LNP and SM states. The interesting final states then involve jets in
association with missing transverse energy and possibly leptons and photons. 
Only by studying the detailed properties of these objects can one hope to 
discriminate among the models. 

The most convincing way to discriminate between 
supersymmetry and its competitors is to measure the spin of the new 
particles: in the Little Higgs and UED models the new states and their SM 
partners have the same spin\footnote{
Robust discrimination between the Little Higgs and UED would require observing 
or ruling out the excited level-2 and higher KK excitations of the UED model,
absent in the Little Higgs.}, while in SUSY models their spins differ by 1/2. 
Measuring spin at the LHC, however, is notoriously difficult. Almost all
existing proposals rely on the observation that, if the
produced strongly interacting state decays via a cascade chain, angular
correlations between the particles emitted in subsequent steps in the 
cascade carry information about spin. (See 
Refs.~\cite{c1,c2,c3,c4,c5,c6,c7,KM}, as well as
a recent review~\cite{SDreview}.) The availability of cascade decays 
with the right properties for this to work, however, depends on the spectrum 
and couplings of the model, and is by no means guaranteed. Moreover, a large 
amount of data is typically needed to alleviate combinatoric and other 
backgrounds.

If unambiguous spin measurments are unavailable, the experiments can still 
attempt a more modest task of model discrimination, {\it i.e.} determining 
which of two or more specific theoretical models provides a better fit to 
the available data. Unlike a direct spin measurement, which would rule
out the entire {\it class} of models with the wrong spin assignment, this 
approach can only discriminate between {\it specific} models. For 
example, if it is found that the minimal Littlest Higgs with T-parity (LHT) 
model cannot fit the data, it does not exclude the possibility that another 
model of the Little Higgs class could provide a better fit. Still, this 
approach can provide valuable information, and is well worth pursuing at the 
LHC, especially at the early stages. 

The goal of our study is to estimate the prospects for model 
discrimination with the CMS detector. As a test case, we consider a very 
simple scenario: We assume that the {\it only} new physics process observable 
at the LHC is pair-production of new color-triplet particles, followed by
their decay into a quark and an LNP. This process occurs at the LHC at a 
significant rate over large parts of the parameter space of the MSSM, LHT, and
UED models. Its detector signature is two hard jets (plus possibly additional
jets from gluon radiation and showering) and missing transverse energy.
Our assumption that no other signatures are observed allows us to focus
on this channel alone, and to understand in detail the issues important for 
model discrimination. It would be straightforward to repeat our 
exercise with more complicated models for the exotic particle production (e.g. 
including color octet pair-production channels) and decay (e.g. including 
cascade chains involving leptons and/or weak bosons).  

The main motivation for our study comes from
the work of Barr~\cite{Barr}, who showed that the angular distributions of 
leptons from the decay of lepton partners produced directly (via electroweak 
processes) in hadron collisions carry information about the lepton partner 
spin. This example is particularly simple since the lepton partners are
always produced in s-channel quark collisions, and their angular 
distribution in the production frame is almost unambiguously determined by 
their spin. However, its utility is somewhat limited by the small cross 
section of the direct lepton partner process. For quark partners, the cross 
sections are larger, but the production mechanism is more complicated: 
both quark-initiated and gluon-initiated processes need to be included, 
and in both cases there are both s-channel and t-channel diagrams. Still, 
as we will show, the model-dependence of the matrix elements can be 
sufficiently strong to yield observable differences between the models. 
Crucially, the differences cannot be removed by simply varying the free 
parameters of the model with the wrong spin assignments: to demonstrate this,
we performed a scan over the parameter space of the ``untrue'' model.
(A recent model-discrimination study of Hubisz {\it et. al.}~\cite{Hubisz},
which studied a situation similar to our test case, also noted the 
significant differences in jet distributions between models with different 
spins, but was restricted to a single benchmark point in each model's
parameter space. The importance of scanning over parameters in model-discrimination 
studies has been recently emphasized in Refs.~\cite{KM,3body}.) 

The rest of the paper is organized as follows: In section 2, we give
a description of the minimally supersymmetric and Little Higgs models 
used in our study, including input parameters and particle spectra
as well as the relevant production processes and decay chains.
We discuss the dominant standard model backgrounds
and the selection cuts that were applied to isolate signal events. 
We then define our observables and describe the statistical analysis.
Section 3 contains the results of our model scan, including 
exclusion plots for 200 pb$^{-1}$, 500 pb$^{-1}$, 1 fb$^{-1}$ and 
2 fb$^{-1}$ of integrated luminosity. 
We summarize our conclusions in section 4. Finally, the appendix 
includes a list of formulas for error estimates, 
a description of our method to calculate covariances, as well as an
example of the angular distribution of jets at an excluded LHT model
point.

\section{Setup}

\begin{table}[t!]
\begin{center}
\begin{tabular}{c|c|c|c|c|c|c|c|c|c}   
final state & $\tilde{u}_L\tilde{d}_L$ & $\tilde{q}_{L,R}\tilde{q}_{L,R}^*$ 
& $\tilde{u}_L
\tilde{u}_L$&$\tilde{u}_R\tilde{d}_R$ & $\tilde{u}_R\tilde{u}_R$ & 
$\tilde{d}_L \tilde{d}_L$ & $\tilde{u}_L \tilde{u}_R^*$ 
& $\tilde{d}_R \tilde{d}_R$ & $\tilde{u}_L^* \tilde{d}_R$ \\ 
&&&&&& $\tilde{u}_L \tilde{d}_R^*$ & $\tilde{u}_R \tilde{u}_L^*$ & 
$\tilde{d}_R \tilde{d}_L^*$ & $\tilde{u}_R^* \tilde{d}_L$ \\ 
&&&&&& \rule[-1.8ex]{0ex}{0ex} 
$\tilde{u}_R \tilde{d}_L^*$ & & $\tilde{d}_L \tilde{d}_R^*$ & \\ \hline
$\sigma$(fb) \rule{0ex}{2.7ex} & 600 & 400 & 300 & 230 & 170 & 80 & 65 & 40 & 30 
\end{tabular}
\caption{Cross sections of squark pair-production processes at the LHC
at the study point of Eq.~\leqn{pars}. Here $q=u,d,s,c$. Factorization 
and renormalization scales are set to 500 GeV. The CTEQ6L1 parton distribution
functions are used. The matrix elements are evaluated at tree level using
{\tt MadGraph/MadEvent}, and no K-factors are applied.}
\label{tab:susy_xsec}
\end{center}
\end{table}

We will focus on the discrimination between the minimal versions of the 
MSSM and the Littlest Higgs with T-parity (LHT)~\cite{LHT,Low,HM,HMNP}. 
Both models have been extensively studied in the 
literature; for reviews, see Refs.~\cite{mssmrev,lhtrev}. Each model contains
color-triplet massive partners for each SM quark: squarks in the MSSM and
T-odd quarks, or TOQs, in the LHT. Also, each model contains a stable 
weakly-interacting particle: the neutralino of the MSSM and the ``heavy 
photon'' (the T-odd partner of the hypercharge gauge boson) of the 
LHT. We will assume that these are the only particles 
that play a role in the LHC phenomenology; the rest of the new states in each 
model are too heavy to be produced. Note that the two minimal models 
have important differences in their particle content: for example, the 
minimal LHT
does not have a color-octet heavy particle, a counterpart of the gluino; 
while the MSSM does not have a T-even partner of the top quark present in the
LHT~\cite{tplus1,tplus2}. In our scenario, however, neither of these 
particles is observed. This null result does not help with model 
discrimination, since we don't know whether the particles don't exist or 
are simply beyond the LHC reach. Model discrimination must rely on the
observed properties of the produced exotic particles or their decay products.

Our strategy is to simulate a large sample of events corresponding to one
of the models (we will choose the MSSM) with fixed parameters, and treat this
sample as ``data''. The question is then, how well can this data be fitted 
with the alternative model, in this case the LHT? It should be emphasized 
that the predictions of the LHT model are not unique, but depend on the LHT
parameters. So, when fitting data, one should look for the point in the 
LHT parameter space that provides the {\it best fit}. The LHT can be said to
be disfavored by data only to the extent that this best-fit point is
disfavored. 

\subsection{``Data''}
\label{sec:data}

For our case study, we assume that the MSSM is the correct underlying theory,
with the following parameters: 
\beqa
m(\tilde{Q}_L^{1,2}) &=& m(\tilde{u}_R^{1,2}) = m(\tilde{d}_R^{1,2}) = 
500~{\rm GeV}~\,;\CR
m(\tilde{Q}_L^3) &=& m(\tilde{u}_R^3) = m(\tilde{d}_R^3) = 
1~{\rm TeV}~\,;\CR
m(\tilde{L}_L^{1,2,3}) &=&  m(\tilde{e}_R^{1,2,3}) 
= 1~{\rm TeV}\,;~~~
A_{Q,L}^{1,2,3} = 0;\CR
M_1 &=& 100~{\rm GeV}\,;~~~M_2 = 1~{\rm TeV}\,;~~~M_3 = 3~{\rm TeV}\,;\CR
M_A &=& 1~{\rm TeV}\,;~~~\mu = 1~{\rm TeV}\,;~~~\tan\beta=10\,.
\eeqa{pars}
All parameters are defined at 
the weak scale, and no unification or other high-scale inputs are assumed. 
The parameter choices are driven by the desire to study a point with very
simple collider phenomenology: the new physics production at the LHC is 
completely dominated by pair-production of the first two generations of 
squarks. The total squark pair-production cross section is 5.0 pb.
Table~\ref{tab:susy_xsec} lists the 22 leading squark pair-production 
processes, which together account for over 98\% of the total. Associated 
squark-gluino production is strongly suppressed by the high gluino mass; the 
cross section (summed over squark flavors) is only 11 fb. Associated
squark-neutralino production is larger, with the total cross section of
about 290 fb. However, these events have only a single hard jet, and will
not pass the analysis cuts (see section~\ref{sec:cuts}). Production of 
third
generation squarks is also strongly suppressed, with a total cross section
of only 17 fb. Thus, in our analysis we will simulate the processes listed in
Table~\ref{tab:susy_xsec}, and ignore all other SUSY production channels. 

Another simplification that occurs at the chosen parameter point is in the 
decay pattern of the produced squarks: they decay into quarks and 
the lightest neutralino (essentially a bino) with a 100\%
probability. This means that in this model, the only place where strong 
evidence for new physics would show up at the LHC is the two jets+missing 
energy channel. We will limit our study to this channel.

\begin{figure}[t!]
\begin{center}
\includegraphics[width=7cm]{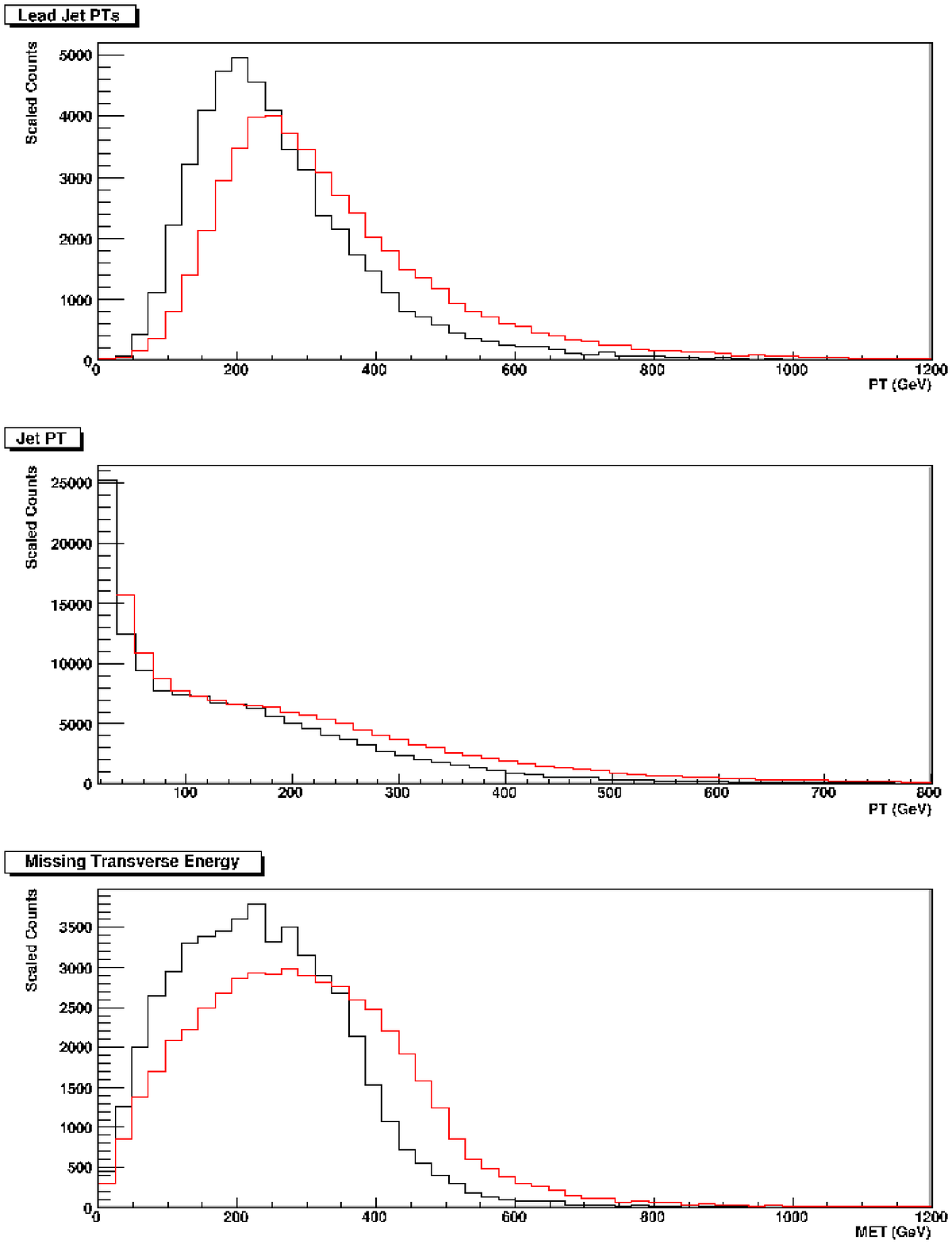}
\hskip1cm
\includegraphics[width=7cm]{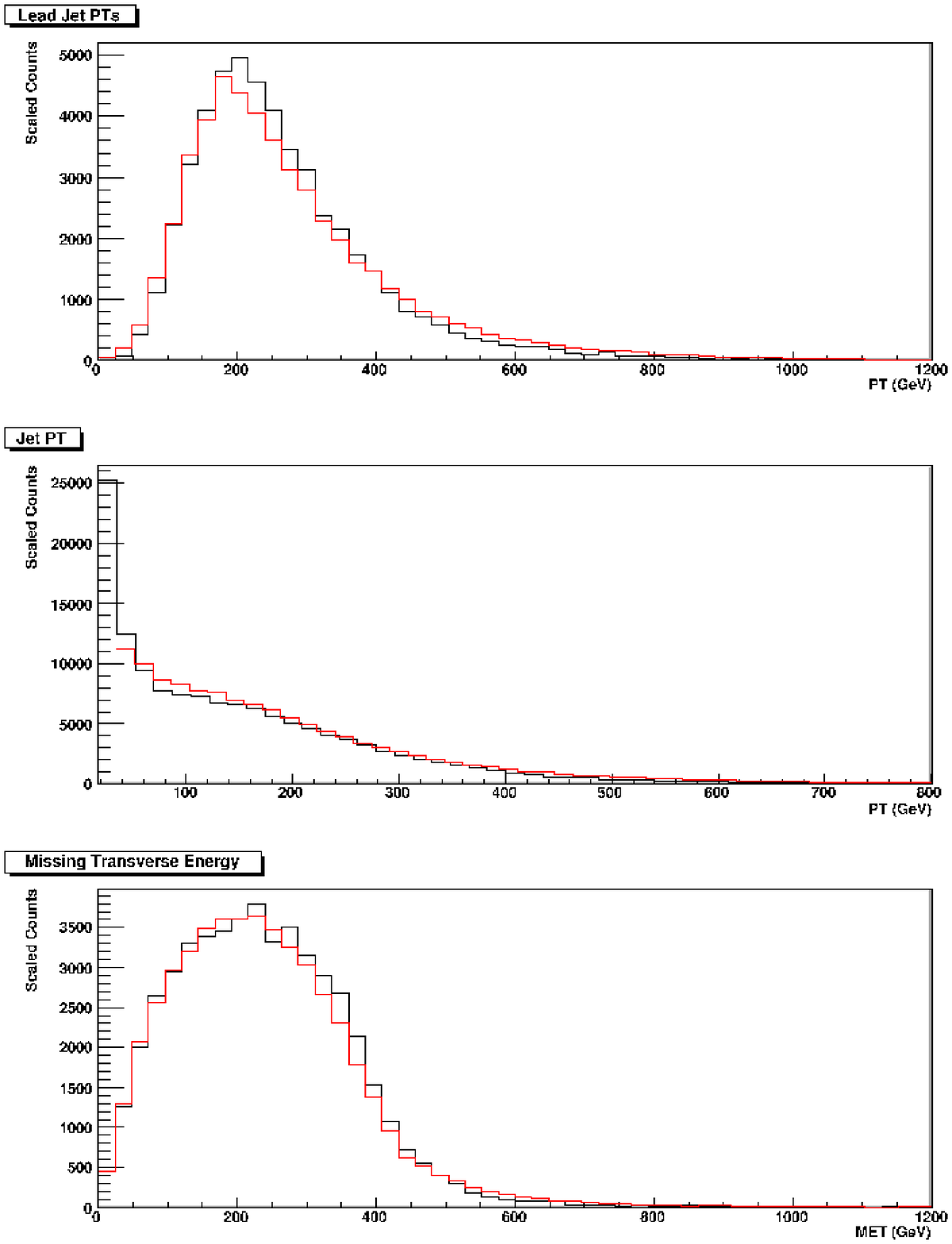}
\vskip2mm
\caption{Jet $p_T$ and missing transverse energy distributions in the MSSM, obtained 
with {\tt PGS} (red histograms) and full CMS simulation (black histograms). Left panel:
Uncorrected {\tt PGS}. Right panel: A jet energy scale correction factor has been 
applied to the {\tt PGS} output.}
\label{fig:PGSvsCMS}
\end{center}
\end{figure}

The ``data'' event sample has been generated in the following way. 
First, we simulate a sample of parton-level events using the 
{\tt MadGraph/MadEvent} package~\cite{MG}.
The production processes included in this simulation, and their cross sections,
are listed in Table~\ref{tab:susy_xsec}. 
The squark decays are also handled by {\tt MadGraph/MadEvent}, using
the narrow-width approximation. The sample size corresponds to 10 fb$^{-1}$
of integrated luminosity at the LHC. The resulting events are stored in a 
format compatible with the Les Houches accord, and then passed on to 
{\tt PYTHIA}~\cite{Pythia} to simulate showering and hadronization. The 
{\tt PYTHIA} output is then passed on to the detector simulation code. 
We use a modified version of the {\tt PGS} code to perform fast (parametrized)
detector simulation. The drastic speed-up of the event simulation provided by 
{\tt PGS} (compared to full CMS 
detector simulation) allows us to scan the 
LHT parameter space, generating a statistically significant event sample 
for each point in the scan. 
To calibrate {\tt PGS} to the CMS detector, 
we have generated two calibration event samples (one in the MSSM and one in
the LHT) using the full CMS detector simulation, and compared them to the
{\tt PGS} output for the same two underlying models. On the basis of this
comparison, we determined that the energy and angular distributions of the {\tt PGS} 
jets are in excellent 
agreement with the full CMS simulation, once the jet energy has been appropriately 
corrected. This is clear from Fig.~\ref{fig:PGSvsCMS}, which shows the jet $p_T$ 
and missing transverse energy distributions in the MSSM, obtained with {\tt PGS} (red 
histograms) and full CMS simulation (black histograms). 
For jets satisfying the selection criteria of our analysis (in 
particular, $p_T^{\rm min}=100$ GeV), the correction factor is essentially the 
same as the one appearing in translation from parton-level jet energy to the 
energy reconstructed by the detector~\cite{CMSnote} (i.e., the {\tt PGS} 
output in this $p_T$ range 
essentially corresponds to parton-level jets). We have applied this
correction factor to the {\tt PGS} output throughout our analysis. 

\subsection{Little Higgs Model}

If the only evidence for new physics is in the two jets+missing energy 
channel, it is natural to try to fit the data with the LHT model, assuming 
the dominant production process
\beq
pp \to \tq_i \bar{\tq}_i\,,
\eeq{proc_lht}
where $\tq_i$ is the TOQ of flavor $i$. We will assume that four flavors
of TOQs, $i=u,d,s,c$, are degenerate at mass $M_Q$ and are within the reach 
of the LHC, with the other two flavors being too heavy to play a role. Once
produced, TOQs promptly decay via
\beq
\tq_i\to q_i B^\prime,~~~\bar{\tq}_i \to \bar{q}_i B^\prime\,,
\eeq{dec_lht}
giving a 2 jets+MET signature. Here $B^\prime$ is the lightest T-odd particle 
(LTP), the heavy photon of mass $M_B$. The LHT predictions in this channel
are sensitive to only two model parameters, $M_Q$ and $M_B$, which 
allows us to scan the parameter space with realistic computing resources.
The counterpart of the process~\leqn{proc_lht},~\leqn{dec_lht} in the $pp$
collisions at the Tevatron was considered in Ref.~\cite{lhttev}. The 
Tevatron experiments exclude a region in the $M_Q-M_B$ plane: roughly,
they place a lower bound of $M_Q\gsim 350$ GeV for light $B^\prime$, 
$M_Q-M_B\gsim 250$ GeV, and somewhat weaker bounds for heavier $B^\prime$.
(There is no bound if $M_Q-M_B\lsim 50$ GeV.) 

To assess how well the data can be fitted with the LHT model, we perform a 
scan in the $(M_Q, M_B)$ plane. We have picked 125 points in the LHT 
parameter space, uniformly scanning in the ranges
\beqa
M_Q &=& [500, 950]~{\rm GeV}\,,\CR
M_B &=& [100~{\rm GeV}\,, M_Q]\,.
\eeqa{ranges}
For each point in the scan, we generate an event sample using the procedure 
outlined in section~\ref{sec:data} above. Each sample corresponds to 10
fb$^{-1}$ of integrated luminosity at the LHC. 

\subsection{Backgrounds} 

\begin{table}[t!]
		\renewcommand{\arraystretch}{1.2}
\begin{center}
\begin{tabular}{lrrrrrrr}
& $\sigma_{tot}$ & $\sigma_1$ &$\sigma_2$ & $\sigma_3$ &
$\sigma_6$ & $\sigma_7$ & $N_{sim}$ \\ \hline \hline
Signal (SUSY) & 5.00 & 4.98 & 4.10 & 2.91 & 2.06 & 0.65 & 10,037\\ \hline
$(Z \to \nu \nu)+jj$ & 271.54 & 259.73 & 94.05 & 64.34 & 10.21 & 0.20 & 543,080 \\
$(W \to \nu \ell)+jj$ & 55.80 & 52.58 & 19.30 & 12.89 & 6.27 & 0.37 & 111,602 \\
$(W \to \nu \tau)+j$ & 138.27 & 92.67 & 12.18 & 2.49 & 0.52 & 0.04 & 276,540 \\
$t\bar{t}$ & 398.52 & 384.14 & 27.85 & 13.89 & 1.62 & 0.04 & 797,039 \\ \hline
total BG & 864.13 & 789.11 & 153.37 & 93.61 & 18.62 & 0.65 & 1,728,261 \\
\end{tabular}
\\[1ex]
\caption{Signal and Background cross sections (in pb), where $\sigma_n$
denotes the cross section after cuts 1 to $n$ (see text for description). 
Also listed are the total number of events simulated for our study.}
\label{tab:xsec}
\end{center}
	\renewcommand{\arraystretch}{1.}
\end{table}

Several Standard Model processes contribute to the jets + missing energy final 
state. The following background processes are dominant:

\begin{itemize}

\item $Z$+2 jets, with $Z$ decaying invisibly (irreducible background);

\item $W$+2 jets, with $W$ decaying leptonically and the charged lepton 
misidentified or undetected;

\item $W$+1 jet, with $W$ decaying to $\tau \nu_\tau$, the $\tau$ 
decaying hadronically and misidentified as a jet

\item $t\bar{t}$, with at least one of the top quarks decaying leptonically and the charged lepton(s) misidentified or undetected.

\end{itemize}

The cross sections for each process are listed in Table~\ref{tab:xsec}. 
(For the $Z/W$+jets channels, we list the parton-level $Z/W+2$ jets cross
sections with $p_T^j\geq 100$ GeV.) We simulated two independent Monte Carlo
samples for each process. One of the samples is mixed with the SUSY events 
to obtain the ``data'' sample, while the other one is mixed with the LHT
events and used to fit the data. The size of each sample corresponds to
2 fb$^{-1}$ of LHC data. 
All samples have been simulated following the same simulation
path as for the signal: parton-level simulation with {\tt MadGraph/MadEvent},
followed by showering and hadronization simulation with {\tt PYTHIA} and
a parametrized detector simulation with the modified {\tt PGS}. It should be kept in 
mind that some of the CMS detector performance parameters which affect the background
rates, such as lepton misidentification probabilities, may not be realistically
modeled by {\tt PGS}. In principle one could normalize these parameters using full CMS
detector simulation, as we did for the jet-energy corrections. However, given 
the preliminary nature of our study, we did not attempt such normalization.  

In addition to the processes listed above, pure QCD 
multi-jet events with mismeasured jets leading to apparent missing energy 
are expected to make an important contribution to the background. However,
until the detector is calibrated with real data, it is difficult to predict
this background. We have not included it in this preliminary analysis. 

\subsection{Triggers and Selection Cuts}
\label{sec:cuts}

Throughout the analysis, we impose the following cuts:
\begin{enumerate}
  \item At least two reconstructed jets in the event
  \item $p^T(j_1) \geq 150~{\rm GeV}$
  \item $p^T(j_2)\geq 100~{\rm GeV}$
  \item $\eta(j_1) \leq 1.7$
  \item $\eta(j_2) \leq 1.7$
  \item No identified leptons ($e$,$\mu$ or $\tau$) in the event
  \item $\met \geq 300~{\rm GeV}$
\end{enumerate}
where the jets are labeled according to their $p_T$, in descending order.
We do not impose any explicit cuts on jet seperation, since
jet reconstruction in PGS effectively acts as a minimum separation cut.
The LHC data samples will correspond to certain trigger
paths, in our case to the \met~trigger and to jet triggers.  Using simple parametrizations for
the trigger efficiencies~\cite{CMS_trigger} we expect them to be essentially 100\% efficient, given our selection cuts.

The signal and 
background cross sections passing each of the selection cuts are listed in Table~\ref{tab:xsec}. After the cuts are 
applied, we obtain
\beq
S/B = 1.0,~~~~~S/\sqrt{B} = 36~~~(2~{\rm fb}^{-1})
\eeq{StoB}
for the SUSY signal. The $S/B$ value is not as good as those obtained in some existing studies of SUSY search prospects (see, for example, 
Refs.~\cite{CMS_TDR,Hubisz}). The reason is that in those analyses 
gluinos are assumed to be light, around 500 GeV, which greatly increases the signal cross section and also yields three or more hard jets in the final state in most events, allowing to further suppress the background. Still, the relatively large new physics cross section implies that if reasonably accurate predictions of the background rate are 
available, the presence of new physics can be convincingly established.    
In particular, using the 10 observables listed below and the assumptions about
the systematic and statistical errors described in Appendix~\ref{app:Error}, we estimate
that the existence of new physics in this channel will be established at 
the level of 2.5, 4.2, and 4.9 sigma, with analyzed data samples of 200 pb$^{-1}$,
500 pb$^{-1}$, and 1 fb$^{-1}$, respectively. The discovery is dominated by shape
observables: if the total rate, which may suffer from large uncertainties in the MC 
predictions, is removed from the fit completely, the confidence levels are only 
marginally lower.

\subsection{Observables}
\label{obs}

Our analysis uses the following observables: 

\begin{itemize}

\item $\sigma_{\rm eff}$: The cross section, in pb, of events that pass the 
analysis cuts. Experimentally, this quantity is inferred from the measured 
event rate using $N_{\rm obs} = {\cal L}_{\rm int} \sigma_{\rm eff}$, 
where $N_{\rm obs}$ is the number of events passing the cuts in a
sample collected with integrated luminosity ${\cal L}_{\rm int}$. 
It is related to the total production cross section by
$\sigma_{\rm eff} = \sum_i \sigma_i E_i$, where the sum is over all 
processes (signal and background) which contribute to the sample, and 
$\sigma_i$ and $E_i$ are the total cross section and combined trigger/cuts 
efficiency, respectively, for channel $i$.

\item $\langle p_T \rangle$: The average transverse momentum of all jets 
with $p_T>100$ GeV in a given data sample that pass the 
analysis cuts. This variable is tightly
correlated with the mass difference between TOQ and the LTP, $M_Q-M_B$.

\item $\langle \; | \Sigma \eta|\; \rangle$: The average of the
absolute value of the sum of the pseudo-rapidities of the two leading 
(highest-$p_T$) jets in the event.

\item $\langle H_T \rangle$: The average of the 
scalar sum of the transverse momenta of all 
jets in the event plus the missing transverse energy
\[ H_T=\sum_{\mbox{\scriptsize{jets}}} p_T^{(i)} + \met . \]

\item $\langle \met \rangle$, the average of the missing 
transverse momentum in the events that pass the selection cuts.

\item {\it Beam Line Asymmetry (BLA):} This observable is defined as 
$(N_+-N_-)/(N_++N_-)$, where $N_+$ and $N_-$ are the numbers of events 
with $\eta_1\eta_2>0$ and $\eta_1\eta_2<0$.  

\item {\it Directional Asymmetry (DA):} The same as above, where 
$N_+$ ($N_-$) are now the numbers of events where $\vec{p}_1 \cdot 
\vec{p}_2$ is positive (negative).\footnote{For a recent analysis using BLA and DA in a context similar to ours, see Ref.~\cite{MR}}. A plot showing
the distribution of relative angles between the two hardest jets 
can be found in Fig.\ref{fig:dadist} in the appendix.

\item {\it Transverse momentum asymmetry (PTA):} The ratio $N^+/N^-$ 
of the number of jets with $p_T$ larger than $\langle p_T \rangle$
and the number of jets with $p_T$ smaller than $\langle p_T \rangle$.

\item {\it Transverse momentum bin ratios:} We distribute the jets
in the event into three fixed bins, depending on their transverse 
momentum. The first bin corresponds to 100 GeV$<p_T<$300 GeV
($N_1$ events),
the second to 300 GeV$<p_T<$500 GeV ($N_2$ events), and the third to 
$p_T>$500 GeV ($N_3$ events). We then define two bin count ratios,
$R_1=N_2/N_1$ and $R_2=N_3/N_1$.

\end{itemize}

\begin{table}[t!]
		\renewcommand{\arraystretch}{1.2}
\begin{center}
\begin{tabular}{l|rrrrrrrrr}
& $\langle p_T \rangle$ & $\langle H_T \rangle$ & $\langle \met \rangle$
& $\langle | \Sigma \eta | \rangle$ & BLA & DA & PTA & R$_1$ & R$_2$ \\ \hline
$\langle p_T \rangle$ & 1 & 0.86 & 0.42 & -0.08 & -0.03 & -0.37 & 0.30 & 0.88 & 0.00 \\ 
$\langle H_T \rangle$ & 0.86 & 1 & 0.66 & -0.10 & -0.05 & -0.34 & 0.22 & 0.76 & -0.06 \\ 
$\langle \met \rangle$ & 0.42 & 0.66 & 1 & -0.04 & -0.04 & -0.06 & 0.05 & 0.35 & -0.11 \\ 
$\langle | \Sigma \eta | \rangle$ & -0.08 & -0.10 & -0.04 & 1 & 0.64 & 0.50 & -0.01 & -0.07 & 0.02 \\ 
$BLA$ & -0.03 & -0.05 & -0.04 & 0.64 & 1 & 0.41 & -0.02 & -0.02 & -0.00 \\ 
$DA$ & -0.37 & -0.34 & -0.06 & 0.50 & 0.41 & 1 & -0.21 & -0.38 & -0.16 \\ 
$PTA$ & 0.30 & 0.22 & 0.05 & -0.01 & -0.02 & -0.21 & 1 & 0.22 & 0.64 \\
$R_1$ & 0.88 & 0.76 & 0.35 & -0.07 & -0.02 & -0.38 & 0.22 & 1 & 0.14 \\
$R_2$ & 0.00 & -0.06 & -0.11 & 0.02 & -0.00 & -0.16 & 0.64 & 0.14 & 1
\end{tabular}
\\[1ex]
\caption{Correlation matrix of observables in the SUSY plus SM background 
``data'' sample, generated from 2 fb$^{-1}$ of simulated events with 
50 subsamples and 10,000 iterations. A description of the procedure 
used to calculate this matrix can be found in appendix \ref{app:Error}.}
\label{tab:corr}
\end{center}
 \renewcommand{\arraystretch}{1.}
\end{table}

We compute the ``measured'' values of these observables using the 
``data'' sample. For each LHT point in the scan, we compute the 
expected central values of the observables using the corresponding MC sample. 
We then use the standard $\chi^2$ technique to estimate the quality of the fit between the expected and measured values. The observables are assumed to be Gaussian distributed, with the variances including 
statistical and systematic errors added in quadruture. The correlation matrix between observables for each LHT point is obtained from the generated Monte Carlo sample; the details of the procedure and 
error analysis are described in Appendix~\ref{app:Error}. As an example, 
we show the correlation matrix for our susy ``data'' sample in 
Table~\ref{tab:corr}.
The quality of
the fit to data at each LHT point is quantified by the $\chi^2$ value, which can in turn be converted into probability that the observed disagreement between the measured and expected values of the observables is the result of a random fluctuation. (If this probability is close to one, the fit is perfect; if it approaches zero, the fit is very poor.) 
As a sanity check to validate our statistical procedure, we simulated a large number of 
independent subsamples of SUSY and SM background events, and confirmed that the distribution
of $\chi^2$ values agrees with statistical fluctuations.

\begin{figure}[t!]
\begin{center}
\includegraphics[width=6cm]{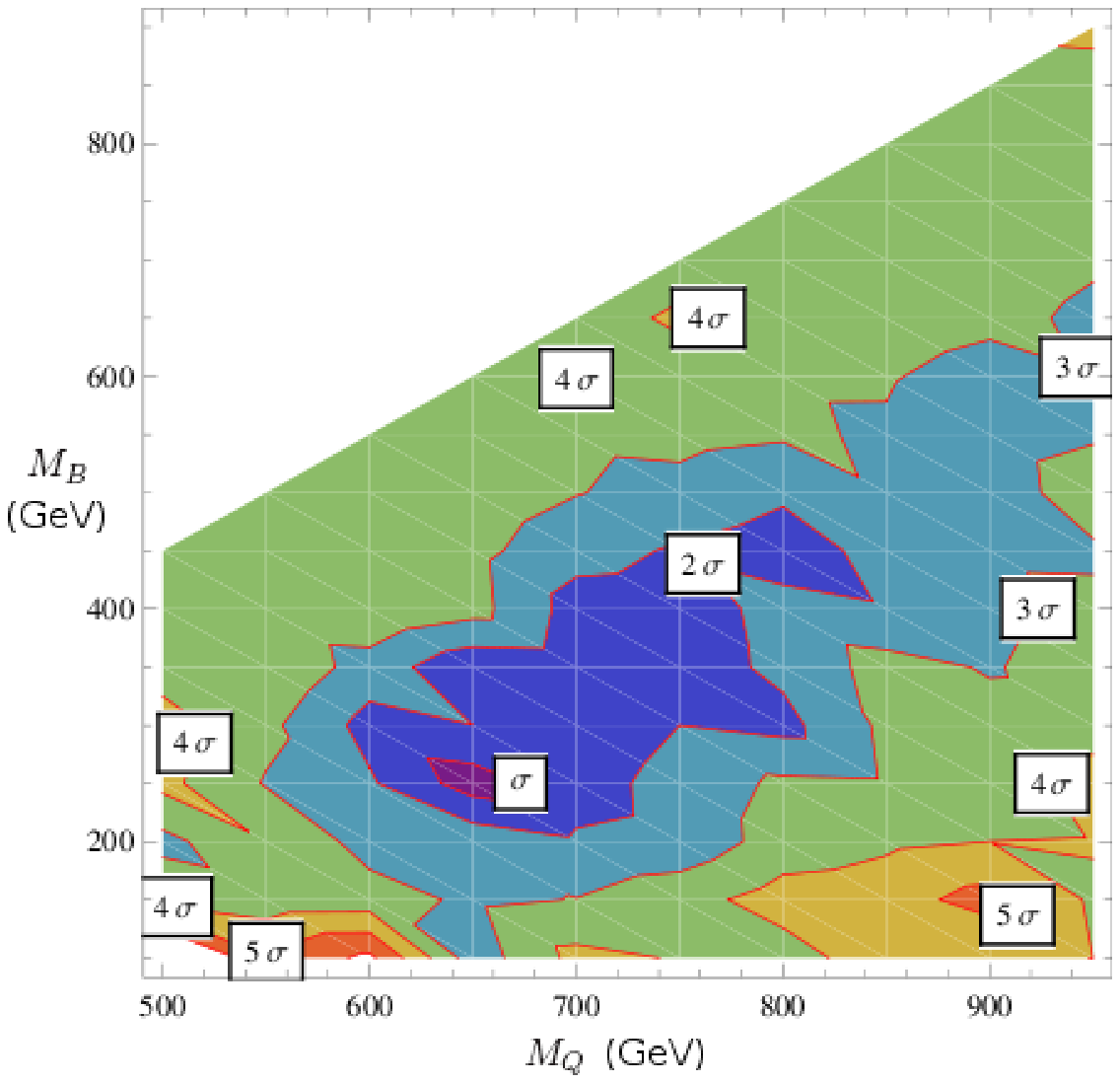}
\hskip1cm
\includegraphics[width=6cm]{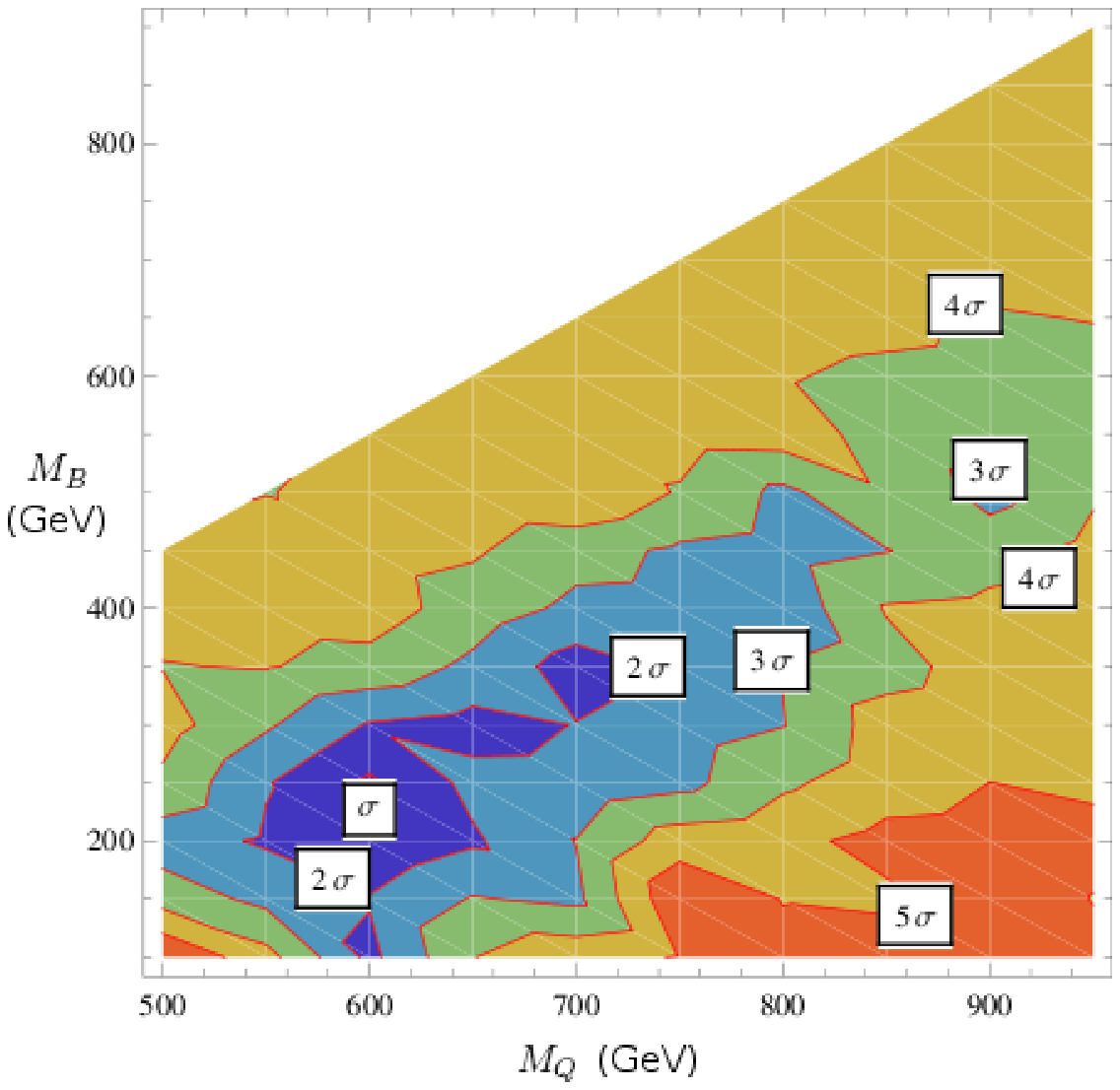}
\vskip2mm
\includegraphics[width=6cm]{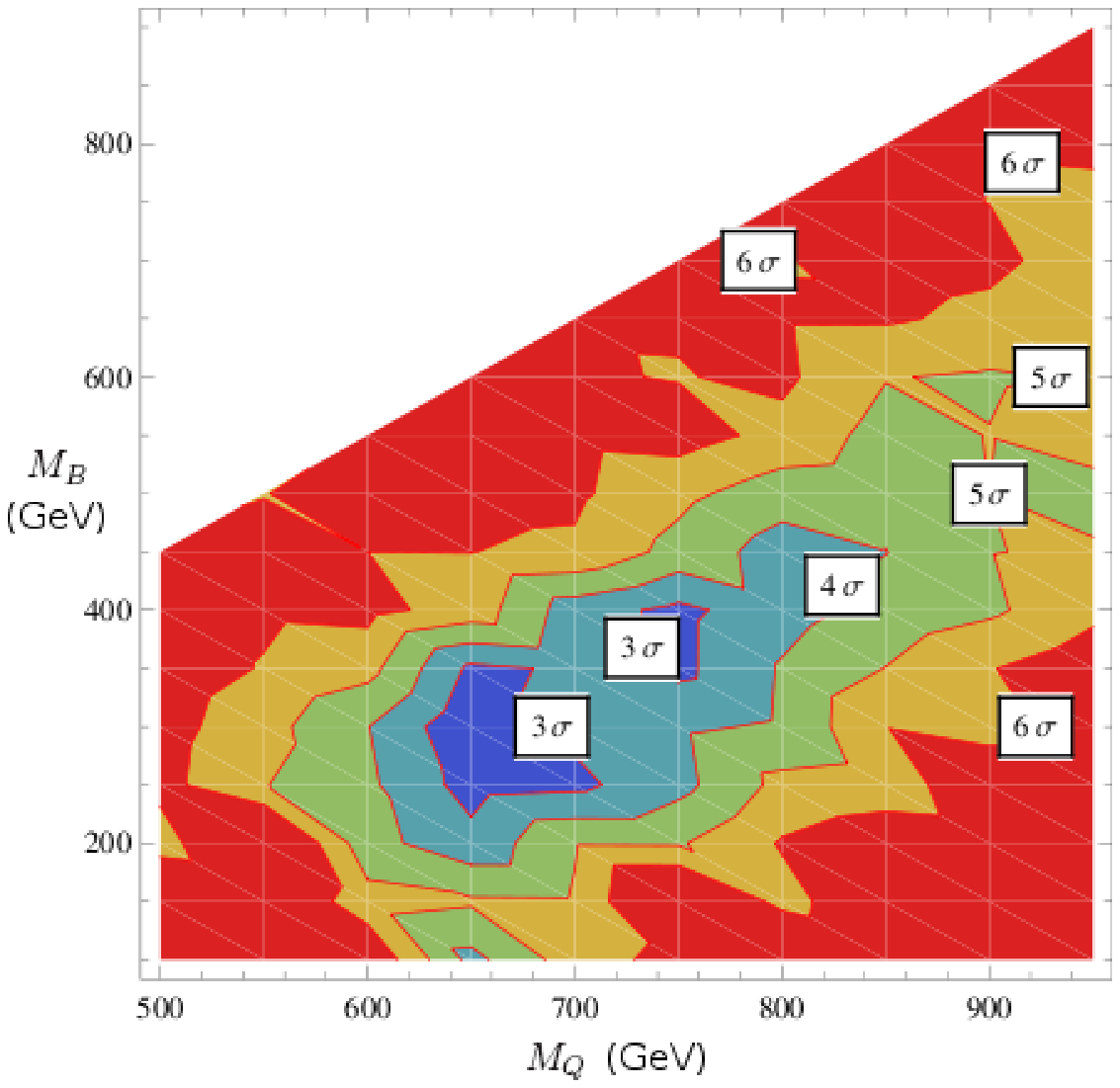}
\hskip1cm
\includegraphics[width=6cm]{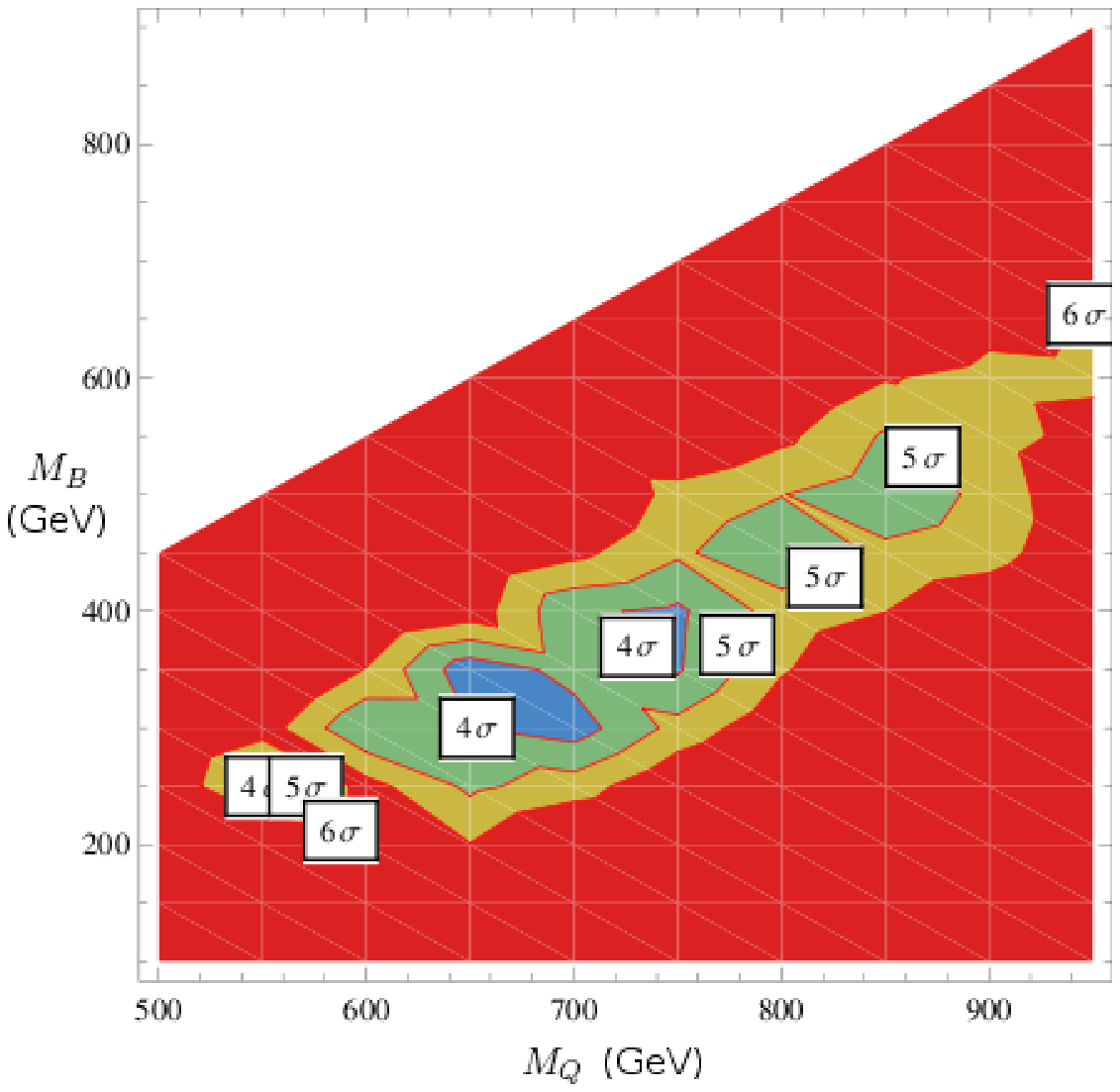}
\vskip2mm
\caption{Exclusion level of the LHT hypothesis, based on the combined fit to the ten observables 
discussed in the text. Top left panel: with integrated luminosity of 200 pb$^{-1}$ at the LHC. 
Top right panel: same, with integrated luminosity of 500 pb$^{-1}$, Bottom left panel: integrated 
luminosity of 1 fb$^{-1}$, Bottom right panel: 2 fb$^{-1}$.}
\label{fig:10Comb}
\end{center}
\end{figure}

\section{Results}

The main results of the analysis are presented in Fig.~\ref{fig:10Comb}, which shows the level 
at which the LHT model is excluded depending on the assumed values of the parameters. For 
illustration purposes, we label the exclusion contours by the number of standard deviations 
in a single-variable Gaussian distribution corresponding to the same probability. With 200 
pb$^{-1}$ of accumulated data, the combined fit to the 10 observables excludes only about 
half of the LHT parameter space at better than 3-sigma level, or better than 99.7\% 
confidence level. In the rest of the parameter space the LHT model is still consistent 
with data at this level, with the best-fit point at $M_Q=650$ GeV, $M_B=250$ GeV showing 
a less than 1-sigma deviation from the data. With more integrated luminosity and 
correspondingly smaller statistical errors, however, the LHT model can no longer fit 
the data. For 2 fb$^{-1}$, we find that the complete LHT parameter space in our study 
is excluded at a more than 3-sigma level, and most of the parameter space is already 
excluded at a 5-sigma level. Thus, it appears that in our test-case scenario, experiment 
can exclude the LHT interpretation of the data with a modest integrated luminosity of 
only a few fb$^{-1}$.

\begin{figure}[t!]
\begin{center}
\includegraphics[width=6cm]{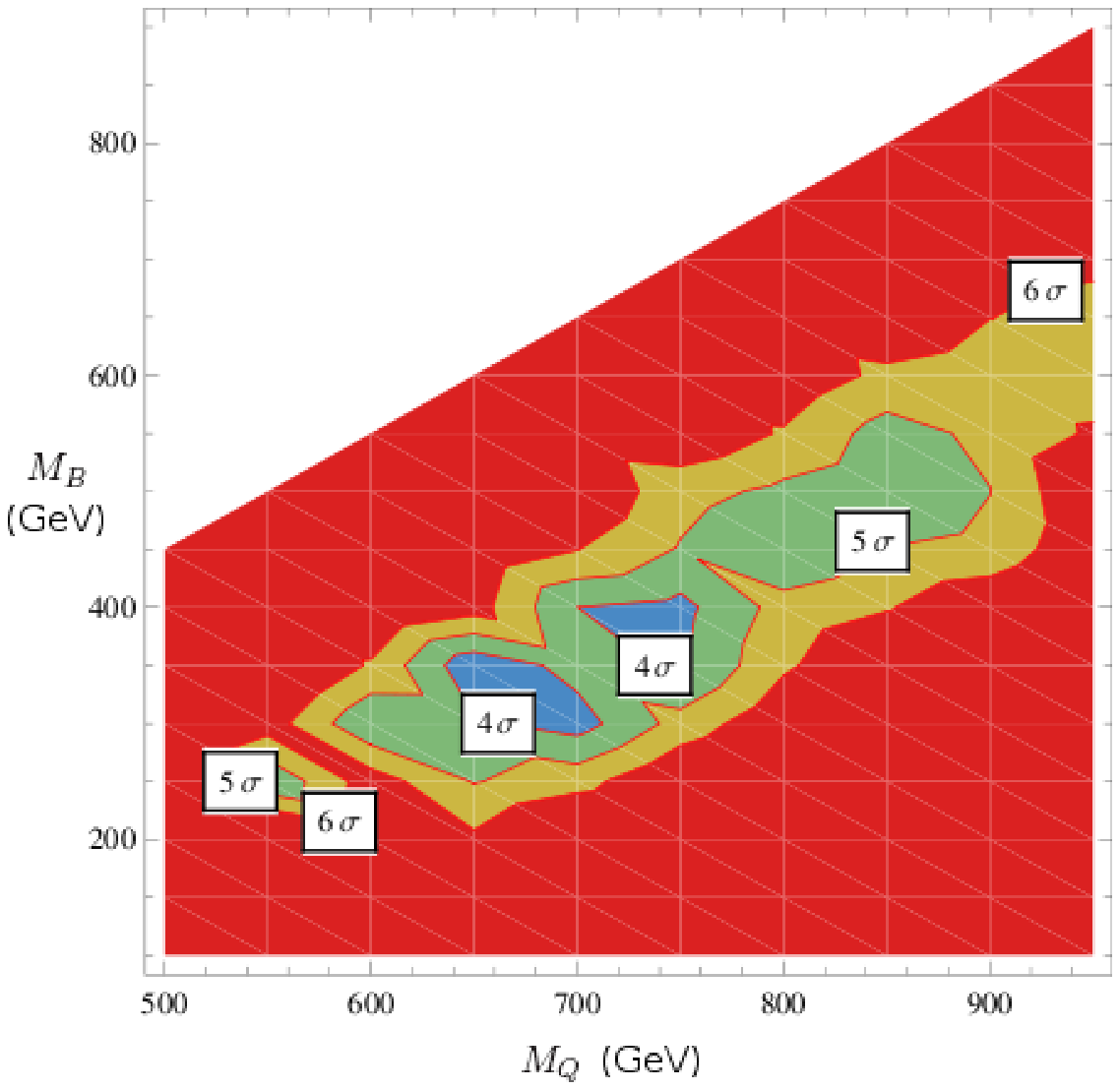}
\hskip1cm
\includegraphics[width=6cm]{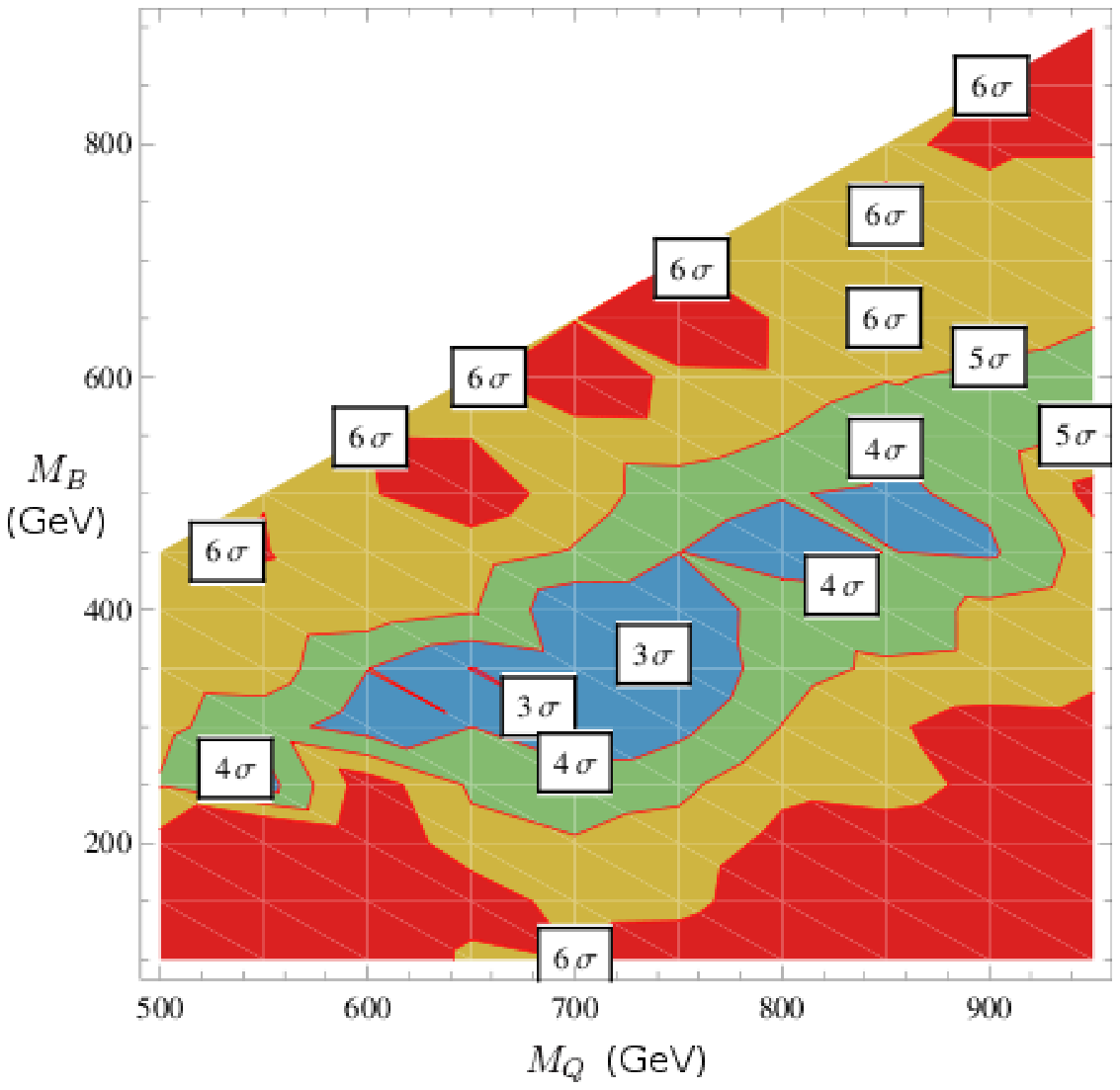}
\vskip2mm
\caption{Exclusion level of the LHT hypothesis, based on the combined fit to 
nine/eight of the ten observables discussed in the text, with integrated luminosity 
of 2 fb$^{-1}$ at the LHC. Omitted are the total production cross section (left panel) 
and missing transverse momentum and $H_T$ (right panel).}
\label{fig:9Comb}
\end{center}
\end{figure}

While we include the estimates of the systematic uncertainties for all observables in our 
study, some of the observables may suffer from additional uncertainties. One example is the 
total production cross section. We assumed a 30\% systematic error on the value of the cross 
section computed in the LHT model, to account for the scale uncertainty of the leading-order 
calculation, as well as pdf and luminosity uncertainties. However, other effects, for 
example the possibility that the number of degenerate TOQ flavors is different from the 
assumed value (four), the possible presence of additional TOQ decay channels, etc., could 
significantly change this observable, keeping all others intact. Thus, it is interesting 
to fit the data with the LHT model without using the cross section information at all. 
Interestingly, this fit leads to exclusion of the LHT model at levels not much weaker 
than the original fit, see Fig.~\ref{fig:9Comb}. In other words, the cross section 
information does {\it not} seem to play a crucial role in model discrimination: a 
combination of transverse-momentum and angular distributions of the two jets is 
sufficient. This is certainly reassuring.  
We have also performed a fit without using the average missing transverse momentum and 
$H_T$ observables, which may suffer from unexpected instrumental systematics.
The results are shown in the right panel of Fig.~\ref{fig:9Comb}. The impact of removing 
these observables is more significant; some parameter values in the LHT model 
are now no longer excluded at the 3-sigma level. If those two observables are not included, 
it would therefore be necessary to increase the integrated luminosity to arrive at the same
confidence level for the rejection of the Little Higgs hypothesis.

\section{Conclusions}

Using Monte Carlo samples, we determined $\chi^2$ values for
fitting a SUSY + BG ``data'' sample with Little Higgs model predictions, 
using the heavy TOQ and ``heavy photon'' masses as fit parameters
and including dominant standard model backgrounds.

With 2 fb$^{-1}$ of signal and background 
events, we were able to show that a combination of ten observables encoding 
angular and transverse momentum distributions of the observed jets 
contains enough information to exclude the LHT model at a 
3-sigma confidence level, provided that these distributions 
for both models and the dominant Standard Model backgrounds 
can be reliably predicted by Monte Carlo simulations.
We found that neither the effective cross section, which
depends on potentially unknown decay branching ratios, 
nor information about the missing energy is crucial for this method of 
model discrimination.

In reality, it is likely that the LHC phenomenology is much richer than the
simple scenario described here, involving, for example, competing SUSY 
production processes 
and complicated decay chains. In this case, the model-discrimination analysis 
would involve multiple channels, and more new particles (and hence 
parameters) 
would be required to fit. However, while the details are highly 
model-dependent, it should be conceptually straightforward to extend our 
analysis to such situations.\\[1ex]

\noindent{\large \bf Acknowledgments} 

The authors acknowledge the CMS collaboration for the permission to use the
detector simulation software, and for useful discussions within the CMS SUSY
physics group.
We are grateful to Jay Hubisz and Frank Paige for useful discussions. This 
research is supported by the NSF grants PHY-0355005, 
PHY-0757894 and PHY-0645484. J.T.~acknowledges the
Cornell University College of Arts and Sciences for their support.

\begin{appendix}

\section{Error estimates \label{app:Error}}

Our estimates for the significance level of model exclusion rely on
correct evaluation of statistical and systematic errors. We 
therefore include a summary of formulas used in our analysis. 
We use three fundamentally different types of
observables. The first class consists of averages of measured
quantities like the mean jet $p_t$ and the mean $H_t$ of events.
Secondly, asymmetries in event shapes, as well as
 bin ratios, are obtained by counting
events that do or do not fulfill certain conditions. Finally, the cross
section is calculated from the total number of signal and background
events after cuts.

\subsection{Mean value observables}

The error in transverse momentum $p_T$ of individual jets is estimated using
the parameterization given in \cite{CMS_TDR}.
\[ 
\sigma_{p_T}=\left( \frac{5.6}{p_T^{\mbox{\scriptsize PGS}}} + \frac{1.25}{\sqrt{p_T^{\mbox{\scriptsize PGS}}}} + 0.033 \right) \, p_T^{\mbox{\scriptsize meas}} 
\]
where all momenta are in GeV, $p_T^{\mbox{\scriptsize{PGS}}}$ is the transverse momentum 
obtained from PGS, and $p_T^{\mbox{\scriptsize{meas}}}$ is the rescaled momentum
as in \cite{CMSnote}.

The average transverse momentum observable $\langle p_T \rangle$ 
is calculated by taking the mean of all jets with a minimum 
$p_T$ of 100 GeV in the events that pass our selection cuts.

The missing transverse energy as given by PGS has to be corrected
to account for the change in jet energy scales. The modified missing 
transverse energy is
\[ \met_{\mbox{\scriptsize{meas}}} = \met_{\mbox{\scriptsize{PGS}}} + \sum_{i=1}^{N_{\rm jets}} 
\left( p_T^{\mbox{\scriptsize{PGS}}} - p_T^{\mbox{\scriptsize{meas}}} \right), \]
where the sum is a vector sum in the transverse plane.

The error in the missing transverse energy $\met$ is estimated as
\[ \sigma_{\met}^2 = (3.8 \mbox{ GeV})^2 + 0.97^2 \mbox{ GeV} \, \met
+ ( 0.012 \, \met )^2 \]
as given in \cite{CMS_TDR}.

The observable $\langle H_T \rangle$ is given by the scalar sum of the transverse momentum
of all objects plus the missing energy in the event. The error of this
quantity is calculated by adding the errors of each object and the missing
energy in quadrature.

Given a list of individual measurements of the jet $p_T$, $\met$, or 
$H_T$, the statistical error of the mean value is given by
\[ \sigma_{\mbox{\scriptsize{stat}}}^2 = V/N , \]
where $V$ is the variance of the distribution and $N$ is the number of 
entries. The systematic error is given by
\[ \sigma_{\mbox{\scriptsize{syst}}}^2 = \nu^2/N, \]
where $\nu$ is the mean value of the distribution.

For the average sum of leading jet pseudorapidities 
$\langle | \Sigma \eta | \rangle$
the statistical error is estimated as above, while the systematic error is 
calculated from
\[
\sigma_{\mbox{\scriptsize{syst}}}^2= \frac{w_c^2}{2N},
\]
where the $\eta$ cell width $w_c$ is 0.087.

\subsection{Counting type observables}

Given two bins of events $N^+$ and $N^-$, 
we define the asymmetry as 
\[ A = \frac{N^+-N^-}{N^++N^-}. \]
We assume purely statistical errors given by 
\[ \sigma^+=\sqrt{N^+}, \quad \sigma^-=\sqrt{N^-}, \]
which leads to an asymmetry error of
\[ \sigma_A^2 = \frac{4 N^+ N^-}{\left(N^++N^-\right)^3}. \] 
Given the same two bins, we define the event ratio as 
\[ R = \frac{N^+}{N^-}. \]
The statistical error is then given by 
\[ \sigma_R^2 = \frac{N^+ N^- + (N^+)^2}{(N^-)^3}. \]

\subsection{Cross Section}

The cross section after cuts $\sigma_{\rm eff}$ is given by 
\[ \sigma_{\rm eff} = N_{\rm obs}/\mathcal{L}_{\rm int}, \]
where $\mathcal{L}_{\rm int}$ is the integrated luminosity and 
$N_{\rm obs}$ the observed number of events. The statistical 
error is given by
\[ \sigma_{\mbox{\scriptsize stat}} = \frac{\sigma_{\rm eff}}{\sqrt{N_{\rm obs}}} \]
and the systematic error is estimated as 30 percent,
\[ \sigma_{\mbox{\scriptsize syst}} = 0.3 \, \sigma_{\rm eff} . \] 

\subsection{Covariance Matrix Estimate}

\begin{figure}[t!]
\begin{center}
\includegraphics[width=6cm]{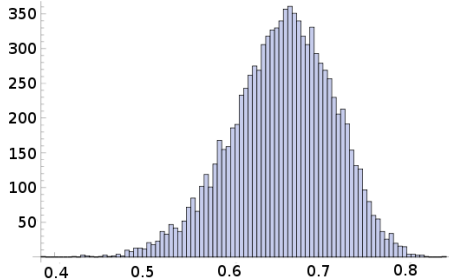}
\caption{Histogram of the $N_{\rm repeat}$ values of the correlation 
between $\langle H_T \rangle$ and $\langle \met \rangle$ obtained 
by applying the bootstrapping procedure on 2 fb$^{-1}$ of SUSY plus 
background events, with $N_{\rm sub}=20$ and $N_{\rm repeat}=10,000$.
The mean value of the distribution is 0.66, as given in 
Table~\ref{tab:corr}.}
\label{fig:CorrElement}
\end{center}
\end{figure}

Preserving information about the expected correlations
of observables can considerably increase or decrease
$\chi^2$ values, depending on the relative signs
of observed deviations from the expected mean values.
It is therefore highly desireable to estimate the elements
of the covariance matrix in a consistent way.

Since it is not possible to
calculate the covariances of all observables $\mathcal{O}_i$ 
analytically, we have to rely on an estimate based on a sample 
of Monte Carlo simulations.

In an ideal world, we would simulate a full sample corresponding to
the desired luminosity at each Little Higgs model point 
(including standard model backgrounds) $N_S$ times and 
estimate the covariance matrix from
\[ V_{ab}=\langle \left( \mathcal{O}_a - \langle \mathcal{O}_a \rangle \right) 
\left( \langle \mathcal{O}_b - \langle \mathcal{O}_b \rangle \right) \rangle , \]
where $\langle \rangle$ denotes the mean over the $N_S$ sets.
Because of limited computing resources, this is not feasible
and we have to estimate the correlations from existing subsets of events
for each data point. We use a bootstrapping procedure, where 
we randomly select $N_{\rm sub}$ subsamples from 2 fb$^{-1}$ of 
signal plus background events. We calculate the correlation 
matrix from those subsamples, repeat the procedure $N_R$ 
times and then calculate the mean values of the correlation matrix 
elements,
\[ C_{ab}= \frac{1}{N_R} \, \sum_{i=1}^{N_R}
\frac{V_{ab}^{(i)}}{\sigma_a^{(i)} \sigma_b^{(i)}}, \]
where $V^{(i)}$ and $\sigma^{(i)}$ are the covariance 
and standard deviations obtained from the $N_{\rm sub}$ subsamples
in iteration $i$.
Those average matrix elements are then assumed to be the correct correlations 
of the observables in the full sample. 
A histogram of results obtained by this procedure is shown in 
Fig.~\ref{fig:CorrElement}. 

Finally, we assume that 
the correlation is independent of the sample size,
and extrapolate to find the covariances for the full set of events
\[ V_{ab} = C_{ab} \sigma_a \sigma_b, \]
where the standard deviations for the two observables $\sigma_a$
and $\sigma_b$ are now calculated from the full sample and 
include both statistical and systematic errors.

\begin{figure}[t!]
\begin{center}
\includegraphics[width=6cm]{plots/2000_all_cov.eps}
\hskip1cm
\includegraphics[width=6cm]{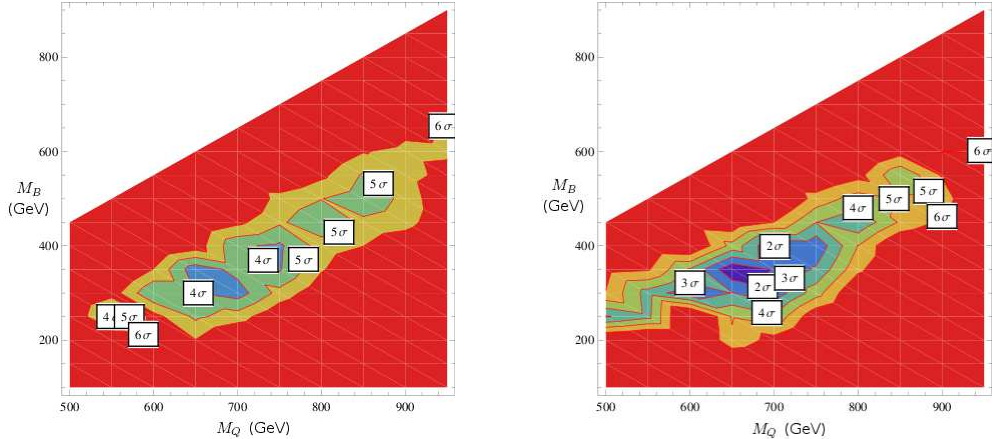}
\vskip2mm
\caption{Exclusion level of the LHT hypothesis, based on the combined fit to the ten observables 
discussed in the text at 2 fb$^{-1}$. Left: As in Fig.~\ref{fig:10Comb}, 
including correlations between observables as determined by the 
bootstrapping procedure. 
Right: Assuming that all observables in the Little Higgs
model are uncorrelated.}
\label{fig:corr_diag}
\end{center}
\end{figure}

We verified that this procedure produces the correct $\chi^2$ 
probability distribution function for the model distance between 
subsample and full sample observables. 

Since the selection of subsample events 
that have passed our cuts is randomized, no information about the 
correlation of the cross section with the other observables can be 
obtained by this method, and we assume that the cross section 
is uncorrelated. 

Fig.~\ref{fig:corr_diag} illustrates the importance of including 
correlation information. Assuming uncorrelated observables, 
a small fraction of points in the LHT parameter space are found 
to be excluded at a higher confidence level. However, the exclusion 
level of the best fit point is lowered significantly, and so the LHT model 
can no longer be rejected at the 3-sigma level.

\section{Angular distribution of jets}

As an example, we show the relative angular distribution of the
two hardest jets in the SUSY ``data'' sample and for the 
Little Higgs model with ($m_Q=500$ GeV, $m_B=100$ GeV).
The directional asymmetry is -0.079 $\pm$ 0.019
for the SUSY ``data'' and 0.008 $\pm$ 0.017 for the LHT + BG sample.
Using just this observable, the $\chi^2$ between the two models
is 26.19, so that it would be excluded at the 5-sigma level.

\begin{figure}[t!]
\begin{center}
\includegraphics[width=8cm]{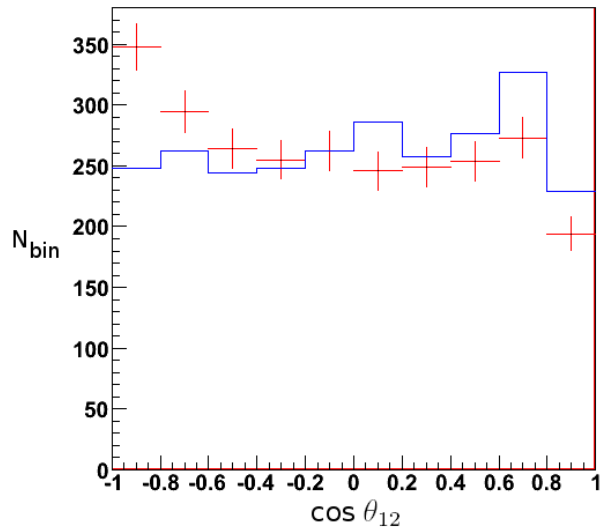}
\vskip2mm
\caption{Distribution of the cosine of the angle $\theta_{12}$ 
between the two hardest jets in the SUSY sample (``data'' points), 
as well as the prediction from the LHT model (histogram) with parameters 
$m_Q=500$ GeV, $m_B=100$ GeV. }
\label{fig:dadist}
\end{center}
\end{figure}

\end{appendix}

\end{document}